# An Ising Hamiltonian Solver using Stochastic Phase-Transition Nano-Oscillators

S. Dutta[1†], A. Khanna[1†], A. S. Assoa[4], H. Paik[3], D. Schlom[3], Z. Toroczkai[2], A. Raychowdhury[4] and S. Datta[1*]

[1]Department of Electrical Engineering, University of Notre Dame, Notre Dame, IN 46556, USA
[2]Department of Physics, University of Notre Dame, Notre Dame, IN 46556, USA
[3]Department of Materials Science and Engineering, Cornell University, Ithaca, NY 14853, USA
[4]School of Electrical and Computer Engineering, Georgia Institute of Technology, Atlanta, GA USA
[†]These authors contributed equally to this work
[*]Corresponding author email: sdatta@nd.edu

**Computationally hard problems, including combinatorial optimization, can be mapped into the problem of finding the ground-state of an Ising Hamiltonian. Building physical systems with collective computational ability and distributed parallel processing capability can accelerate the ground-state search. Here, we present a continuous-time dynamical system (CTDS) approach where the ground-state solution appears as stable points or attractor states of the CTDS. We harness the emergent dynamics of a network of phase-transition nano-oscillators (PTNO) to build an Ising Hamiltonian solver. The hardware fabric comprises of electrically coupled injection-locked stochastic PTNOs with bi-stable phases emulating artificial Ising spins. We demonstrate the ability of the stochastic PTNO-CTDS to progressively find more optimal solution by increasing the strength of the injection-locking signal – akin to performing classical annealing. We demonstrate in silico that the PTNO-CTDS prototype solves a benchmark non-deterministic polynomial time (NP)-hard Max-Cut problem with high probability of success. Using experimentally calibrated numerical simulations and incorporating non-idealities, we investigate the performance of our Ising Hamiltonian solver on dense Max-Cut problems with increasing graph size. We report a high energy-efficiency of $1.3 \times 10^7$ solutions/sec/Watt for 100-node dense Max-cut problems which translates to a 5x improvement over the recently demonstrated memristor-based Hopfield network and several orders of magnitude improvement over other candidates such as CPU and GPU, quantum annealer and photonic Ising solver approaches. Such an energy efficient hardware exhibiting high solution-throughput/Watt can find applications in industrial planning and manufacturing, defense and cyber-security, bioinformatics and drug discovery.**

Combinatorial optimization is ubiquitous in various fields such as artificial intelligence, bioinformatics, drug discovery, cryptography, quantitative finance, operations research, resource allocation, trajectory and route planning. Such problems belong to the NP-hard or NP-complete complexity class, requiring computational resources (time and/or energy) that scale exponentially with the problem size. Interestingly, many combinatorial optimization problems can be translated into another fundamental physics problem of finding the ground state of an Ising model (*1*) (or its equivalent Quadratic Unconstrained Binary Optimization (QUBO) problem (*2*)). The Ising model, describing the property of spin glass, was put forward as a tool of statistical physics to explain the phenomenon of ferromagnetism. The Ising Hamiltonian with $N$ discrete spins $\sigma_{1 \leq i \leq N} \in \{-1, +1\}^N$ and a symmetric coupling matrix $J$ in the absence of an external magnetic field is given by $H = -\sum_{i=1}^{N} J_{ij} \sigma_i \sigma_j$.

Finding the ground state of Ising model belongs to the NP-hard complexity class (*3*) and can be extended to other NP-hard and NP-complete problems including all of Karp's twenty-one NP-complete problems through polynomial time mapping (*4*). Solving the Ising model using exact methods such as the branch-and-bound algorithm are often limited to problem sizes of only a few hundred variables. Alternatively, approximate algorithms or heuristics and stochastic approaches such as semidefinite program (*5*), breakout local search (*6*), metropolis algorithm (*7*) and simulated annealing (*8*) are widely used in digital computers to find an optimal or near-optimal solution. Even for moderately sized problem instances, the time to find a near-optimal solution can become prohibitively large. Hence, there is a growing interest towards finding hardware approaches, beyond digital CMOS, that can solve large-scale constrained optimization problems efficiently. Recently, various schemes for building annealing-inspired non-von Neumann processors, called Ising machines, have been devised on a variety of technology platforms. These include super-conducting qubit-based adiabatic quantum computing (AQC) and quantum annealing (*9, 10*), digital and mixed-signal complementary metal-oxide-semiconductor (CMOS) annealers (*11–13*) and coherent networks of degenerate optical parametric oscillators (*14, 15*). Qubit-based quantum annealers incur high cost and complexity arising from operation in cryogenic environment. The optical coherent Ising machine has shown competitive performance compared to the quantum annealer (*16*), but requires long fiber ring cavity for implementing Ising spins using temporal multiplexing and extremely fast (and power hungry) field-programmable gate array (FPGA) for implementing coupling in a measure-and-feedback scheme (*15*). Digital CMOS annealers (*11–13*) rely on an external source for random number generation for introducing stochasticity and find it technologically challenging to maintain true randomness in CMOS implementation, while requiring significant post-processing.

In this work, we propose and demonstrate an electronic phase-transition nano-oscillator (PTNO)-based Ising Hamiltonian solver utilizing the concept of continuous-time dynamical system (CDTS). Specifically, the "phase" state variable $\theta(t)$ is used to represent the Ising spin and the coupling between the oscillators mimic the Ising interaction. By carefully choosing the coupling matrix, the optimization problem can be encoded onto the PTNO network such that the Ising energy is represented as the internal *"energy"* or the Lyapunov function $E(\theta)$ of the PTNO network and the stable point (or the attractor state) of the network represents the solution of the problem. The continuous-time dynamics of the PTNO network and the evolution of the oscillator phases $\theta(t)$ will then be determined by the inherent energy minimization property of the network and can be described by $\frac{d\theta(t)}{dt} = -\frac{\partial E(\theta)}{\partial \theta}$. A vast repertoire of such emergent dynamics exhibited by dynamical systems has been exploited across a wide range of fields from understanding neural activity (*17–19*) to robotic locomotion control (*20, 21*) to solving optimization problems using discrete-time Hopfield networks (*22, 23*). The main advantage provided by PTNO-based Ising solver is its inherent distributed nature and continuous-time dynamics that allows completely synchronous updates of all the Ising spins. This drastically brings down the time for each anneal cycle to one oscillation time period and consequently the time-to-solution. Additionally, the PTNOs operate at room temperature and consume extremely low power which translates to an extremely competitive energy-efficiency.

**Overview of PTNO-based CTDS as an Ising Hamiltonian solver**

The combinatorial optimization problem is reformulated in terms of the Ising Hamiltonian $H$ defined by the spin vector $\vec{\sigma}$, and the symmetric coupling matrix $J$ and mapped onto the Ising solver as shown in Fig. 1(a) and (b). Each Ising spin, representing a node in the graph is emulated by an insulator-to-metal phase-transition (IMT) nano-oscillator. A PTNO comprises a two-terminal phase-transition hysteretic device in series with a transistor as shown in Fig. 1(c). We use Vanadium Dioxide ($VO_2$) as a prototypical IMT material in our experiments (*24*). A false-colored SEM image of a two-terminal $VO_2$ of length = 100nm used in our experiments is shown in Fig. 1(c). The working principle of $VO_2$-based PTNO has been reported earlier elsewhere (*20, 25, 26*). Below the phase-transition temperature and under zero external electric field or current, $VO_2$ shows insulating behavior. Upon application of an electric field across the two terminals of the device that forces current to flow through, the material undergoes an abrupt phase transition from insulating to metallic state. The hysteretic phase transition is reflected as an abrupt hysteretic current-voltage (I-V) characteristic of the device. Pairing a conventional n-type metal–oxide–semiconductor (NMOS) transistor in series with the $VO_2$ such that the load line passes through the unstable region of the I-V curve results in self-sustained oscillations as show in fig. 1(c). We used $V_{DD} = 2V$ and a gate voltage $V_{GS} = 0.8V$ in our experiments. We create a highly interconnected PTNO-network where the coupling matrix W is derived from the adjacency or coupling matrix $J$ of the Ising model as shown in Fig. 1(a) and (b). In this example of Ising model with antiferromagnetic interactions, the presence of an edge between two nodes is denoted as $J = -1$ and is represented by a coupling capacitance $C_C$.

Fig.1(d) shows an overview of the experimental setup of the PTNO-based CTDS. The main computing kernel comprises eight PTNOs connected using coupling capacitance following the coupling matrix W. The fabricated $VO_2$ device array (labeled as 1-8 in Fig. 1(d)) is connected with eight NMOS transistors in series to create eight PTNOs. To emulate artificial Ising spins $\vec{\sigma}$, an external injection locking signal $S_{INJ}$ is applied to all the oscillators using injection capacitances $C_{INJ}$. As explained later, this creates bi-stable oscillator phases $\vec{\theta}$ that are used as state variables for computing. The continuous-time dynamics of the network is dictated by an *"energy"* or the Lyapunov function $E(\vec{\theta})$ that closely resembles the Ising Hamiltonian $H$. The dynamics of the network evolves continuously in time, so as to naturally minimize $E(\vec{\theta})$, and in the process minimizes the Ising Hamiltonian to reach the ground state as shown in Fig. 1(e). Once the network reaches an energy minima state, the output of the network is read out in terms of the state variables $\vec{\theta}$ and subsequently reformulated to provide the optimal solution of the original optimization problem. One formidable challenge in solving combinatorial optimization problem is the inevitable increase in the complexity of the energy landscape with the problem size. Specifically, the presence of a large number of local minima degrades the probability of reaching a global optimum. Stochasticity is a well-known technique utilized in Boltzmann machines and simulated annealing, to overcome the issue of getting trapped in local minima. In the latter case, a stochastic noise is used to perturb the state of the system and the magnitude of noise is reduced slowly over time (replicated as a change in the temperature parameter in the algorithm) as the system approaches the global optimum. In our system, we exploit the inherent stochasticity present in the IMT material (*27*) to escape the local minima as well as introduce a novel way of gradually reducing the temporal fluctuations in the oscillator phases by increasing the strength of the injection locking signal $S_{INJ}$. This allows us to perform classical annealing operation and obtain progressively better solutions as illustrated schematically in Fig. 1(e).

**Artificial Ising spin**

The binary degrees of freedom in the phase space of PTNOs arises from the phenomenon of second harmonic injection locking. We apply a sinusoidal injection locking signal $S_{inj} = V_{inj} sin(2\pi f_{inj} t)$ with $f_{inj} = 2f_0$ as an external input to the oscillators across the capacitor $C_{inj} = 20\ pF$ as shown in Fig. 1(d). When $V_{inj} = 0$, the oscillator is free running. Fig. 2(a, i) shows the voltage output waveform $V_{out}$ measured over multiple runs. The corresponding phase of the oscillator, measured with respect to a reference sinusoidal signal of same frequency $f_0$, shows a constantly varying phase with a uniform probability distribution over the entire phase space as seen Fig. 2(b, i). In contrast, when the oscillator is perturbed with $S_{inj}$ at the first harmonic, i.e. $f_{inj} \approx f_0$ also referred to as first harmonic injection locking (FHIL), $V_{out}$ shows a constant $80^0$ phase locking configuration with $S_{inj}$ as seen in Fig. 2(a, ii). The probability distribution of the phase, measured over multiple runs, shows a single gaussian distribution as shown in Fig. 2(b, ii). Interestingly, when $f_{inj} \approx 2f_0$, the oscillator waveform shows both in-phase ($40^0$) and out-of-phase ($220^0$) injection locking configuration when measured over multiple runs as seen in Fig. 2(a, iii). The corresponding probability distribution shown in Fig. 2(b, iii) exhibits a double gaussian distribution highlighting an equiprobable and bi-stable phase portrait. This bi-stability provides an ideal means to encode the Ising spin in the electrical domain, where phase $40^0$ represents up-spin, i.e., $= +1$, and phase $220^0$ represents down-spin, i.e., $\sigma = -1$.

The continuous-time dynamics of the phase difference $\theta$ between the oscillator and the injection locking signal can be described by a generalized version of Adler's equation (Gen-Adler) given by

$$\frac{d\theta(t)}{dt} = -(f_{inj} - n_H f_o) + K_{inj}^H \int_0^{2\pi} \xi(\theta(t) + \vartheta) \cos(\vartheta)\, d\vartheta \qquad (1)$$

where $n_H$ is the n[th] harmonic of the IMT nano-oscillator and $K_{inj}^H = 2\pi n_H f_o f_{inj} C_{inj} V_{inj}$. The first term describes the frequency mismatch between $V_{out}$ and $S_{inj}$, which contributes to *phase slipping*. The second term depends on the phase delay incurred due to the perturbation caused by $S_{inj}$ and is described in terms of the perturbation-projection-vector (PPV), $\xi$ (see Supplementary information section S1 and S3 for details). The corresponding *"energy"* function or Lyapunov function of the oscillator is given by

$$E(\theta) = (f_{inj} - n_H f_o)\theta - K_{inj}^H \int_0^\theta \int_0^{2\pi} \xi(\phi + \vartheta) \cos(\vartheta)\, d\vartheta\, d\phi \qquad (2)$$

The first energy term comes from the contribution of the frequency mismatch that creates an overall bias in the energy landscape. However, this being a linear term does not introduce any additional valleys or peaks in the energy landscape. The second term describes the interaction between the injection locking signal and the oscillator. The corresponding probability distribution of the oscillator's phase can be calculated as $P(\theta_i) = \frac{e^{-E(\theta_i)/\eta}}{Z}$, where $Z = \sum_i e^{-E(\theta_i)/\eta}$ is the partition function and $\eta$ is analogous to the $kT$ term in the Boltzmann distribution and can be interpreted as a measure of the stochastic noise in the IMT oscillator. Expectedly, with zero injection locking, the *"energy"* function of the oscillator stays flat as shown in Fig. 2(b) indicating a uniform distribution over the phase space as is experimentally obtained in Fig. 2(a). For first harmonic injection locking (FHIL), as long as the relative frequency difference $(f_{inj} - f_o)$ is small compared to the second term in equation, $\frac{d\theta(t)}{dt} = 0$ exhibits one stable point at $\theta \cong 0.4\pi$, i.e. one injection-locked equilibrium phase and the calculated $E(\theta)$ (assuming zero frequency mismatch) displays a single energy minima. This results in a single gaussian peak in the probability distribution of the oscillator's phase as verified experimentally. In the case of SHIL, $\frac{d\theta(t)}{dt}$ exhibits two stable points at $\theta \cong 0.2\pi$ and $\theta \cong 1.2\pi$, i.e. two equilibrium phases. The calculated $E(\theta)$ (assuming zero frequency mismatch) evolves into a double well energy landscape that results in a double gaussian distribution in the phase space as reproduced faithfully in the measurements (see Supplementary information section S1 and S3 for further details).

The steady-state analysis of the Gen-Adler equation in the case of SHIL also predicts that, in the presence of a frequency mismatch between the oscillator and the injection locking signal, there exists a critical amplitude of the injection signal $V_{inj}$ below which no stable solution exists. In our experiments, this critical amplitude is close to 1V for a frequency mismatch of 0.1% as observed in the measurements. Above this threshold, the bi-stability in the phase space begins to appear. Fig. 2(c) shows the measured oscillator phase as a function of the amplitude of the injection signal, $V_{inj}$. For very low $V_{inj}$ close to the critical value, the phase of the oscillator measured over multiple runs shows a wide distribution in the phase space, with the distribution narrowing as $V_{inj}$ increases. This behavior can be understood by considering the perturbation in the energy landscape as shown in Fig. 2(e) for $V_{inj} = $ 1V, 3V and 5V. For $V_{inj} = 1V$, the energy barrier $E_B$ separating the two stable equilibrium phases is low (around $E_B$ =0.006 calculated from Eq. 3). Hence, in the presence of stochastic noise, the oscillator's phase constantly fluctuates between the two stable phases as seen in the Fig. 2(d). An increase in $V_{inj}$ to 3V and 5V increases the barrier height $E_B$ to 0.012 and 0.02, respectively. This reduces the fluctuations in the measured oscillator's phase. This is reflected in the experimentally measured mean time between each phase flip, referred to as the dwell time $\tau_{Dwell}$ (analogous to *Neel relaxation time* for magnetization) as a function of $V_{inj}$ as shown in Fig. 2(f). The increase in $\tau_{Dwell}$ with increasing $V_{inj}$ accurately follows the Arrhenius's relation $\tau_{Dwell} = \alpha \tau_0 e^{\frac{E_B}{\eta}}$, where $\tau_0 = \frac{1}{f_0}$ is the characteristic or attempt time (equal to the time period of the oscillator), $\alpha$ is the fitting parameter and $\eta$ is the stochastic noise in the IMT oscillator. This characteristic of reduction in the temporal fluctuations of the oscillator's phase with increasing amplitude of injection locking signal proves to be key knob towards performing classical annealing in our hardware.

**Replicating the interaction term in the Ising Hamiltonian**
To implement a PTNO-based Ising solver that can replicate an artificial Ising spin system, the oscillators need to be connected to one another using coupling elements that emulate the ferromagnetic and antiferromagnetic nature of interaction. We first study the nature of interaction in a pairwise coupled oscillator system as shown schematically in Fig. 3(a) using capacitance $C_C$ as the coupling element. Fig. 3(b) shows the experimentally measured phase distribution of the two oscillators using an injection locking capacitance $C_{inj} = 20 \, pF$ and coupling capacitance $C_C = 56 \, pF$. The oscillators remain out-of-phase with each other and the two configurations: ($40^0$, $220^0$) and ($220^0$, $40^0$) are two equally probable states. We measured the probability of out-of-phase configuration for varying coupling strength as shown in Fig. 3(c). The probability remains close to 0.5 for low coupling capacitance, meaning both in-phase and out-of-phase configurations are equally probable, and increases close to 1 for stronger coupling. We also compare our experimental results with experimentally calibrated PPV-based numerical simulations (methodology described in the Materials and Methods section later) as shown in Fig. 3(c) showing excellent agreement. To understand the exact nature of capacitive coupling, we compare it with a 2-spin Ising model with antiferromagnetic interaction (negative $J$) where the individual spins prefer to remain anti-parallel with one another. In such a system, the probability of one of the possible configurations- (↑↑,↓↓,↑↓,↓↑) is determined by the Boltzmann distribution $P(\sigma_1, \sigma_2) = e^{-H(\sigma_1,\sigma_2)/kT}/Z$, where $T$ is temperature and $Z = \sum_{\sigma'_1,\sigma'_2} e^{-H(\sigma'_1,\sigma'_2)/kT}$ is the partition function. With antiferromagnetic interaction (negative $J$), the probability for anti-parallel configuration (↑↓,↓↑) increases upon varying the interaction strength from $J = 0$ to $-2$. Thus, the antiferromagnetic interaction in an Ising Hamiltonian can be faithfully replicated using capacitive coupling in this IMT nano-oscillator-based system.

Mathematically, the continuous-time dynamics of such a PTNO network can be further described by extending Eq. 2 and Eq. 3 to incorporate an additional coupling term and is given by

$$\frac{d\theta_i(t)}{dt} = -(f_{inj} - n_H f_{o,i}) + K^H_{inj,i} \int_0^{2\pi} \xi_i(\theta_i(t) + \vartheta) \cos(\vartheta)\, d\vartheta$$
$$+ f_0 \sum_{j=1, j\neq i}^{N} \int_0^{2\pi} \xi_i(\theta(t) + \vartheta) I_{osc,j}\, d\vartheta \qquad (3)$$

where $I_{osc,j} = C_{C,j} \frac{dV_{osc,j}}{dt}$. The additional third term describes the coupling interaction energy between pairs of oscillators. The corresponding *"total energy"* function or the global Lyapunov function of the PTNO-CTDS is then given by

$$E(\vec{\theta}) = \sum_{i=1}^{N}(f_{inj} - n_H f_0)\theta_i - K^H_{inj}\sum_{i=1}^{N}\int_0^{\theta_i}\int_0^{2\pi}\xi_i(\phi + \vartheta)\cos(\vartheta)\, d\vartheta\, d\phi$$
$$- f_0 \sum_{i,j=1,i\neq j}^{N} \int_0^{\theta_i}\int_0^{2\pi} \xi_i(\phi + \vartheta) I_{osc,j} d\vartheta\, d\phi \qquad (4)$$

We use Eq. 4 to calculate the two-dimensional energy landscape for pairwise capacitively coupled IMT nano-oscillators as shown in Fig. 3(b). The calculated energy landscape exhibits four stable points or attractor states – two degenerate global minima at the out-of-phase configuration and two degenerate local minima at in-phase configuration. By increasing the strength of the capacitive coupling, the attractor states for the out-of-phase configuration becomes more prominent.

A similar investigation is performed for pairwise resistively coupled oscillators as shown schematically in Fig. 3(d). Contrary to the previous case, the measured oscillator phases as well as numerical simulations reveal a higher probability to be in-phase with each other in either ($40^0$,$40^0$) or ($220^0$,$220^0$) configuration as shown in Fig. 3(e) for a coupling resistance of $R_C = 40 k\Omega$. To establish the nature of resistive coupling, the probability of in-phase configuration is measured as a function of varying $R_C$ and compared to a 2-spin Ising model with ferromagnetic interaction (positive $J$). Note that a lower $R_C$ represent a higher coupling strength and hence a higher $J$. The increase in the probability of in-phase configuration with decreasing $R_C$, i.e. increasing coupling strength, agrees well with the theoretical Ising model that predicts an increase in probability of parallel configuration (↑↑, ↓↓) for increasing $J$ as shown in Fig. 3(f). We also use Eq. 4 with $I_{osc,j} = -\frac{V_{osc,j}}{R_{C,j}}$ to calculate the two-dimensional energy landscape as shown in Fig. 3(e), revealing two global and two local energy minima for in-phase and out-of-phase configurations, respectively. This establishes that a ferromagnetic interaction can be replicated using resistive coupling in out PTNO-CTDS based Ising solver. By increasing the strength of the resistive coupling, the attractor states for the in-phase configuration becomes more prominent.

**Experimental Demonstration of Max-Cut and Performance Enhancement with Annealing**
We investigate the performance of the PTNO-CTDS-based Ising Hamiltonian solver on a NP-hard graph problem of Max-Cut for an undirected and unweighted graph. The Max-Cut problem statement is defined as: Given an undirected graph $G = (V, E)$ with $V$ nodes and $E$ non-negative weights on its edges, the problem requires partitioning $G$ into two subsets $W$ and $X$ such that the total weight on the edges connecting the two subsets is maximized. The Max-Cut problem can be formulated into an equivalent Ising problem using antiferromagnetic interaction ($J = -1$) and we assume the linear Zeeman term to be zero. The cut size for a given spin configuration $\vec{\sigma}$ has a direct mapping to the Ising Hamiltonian $H(\vec{\sigma})$, given by $C(\vec{\sigma}) = -\frac{1}{2}\sum_{1\leq i<j\leq N}J_{ij} - \frac{1}{2}H(\vec{\sigma})$. As such, minimizing the Ising Hamiltonian $H$ maximizes the cut-set $C$. We chose an undirected and unweighted Mobius Ladder graph with 8 nodes as shown in Fig. 4(a) for our experiment. The phases of the oscillators $\vec{\theta}$ are converted to the Ising spins $\vec{\sigma}$ using discretization windows in the phase space (see Methods section for detail). The PTNOs are connected following the adjacency (or connectivity) matrix of the given graph $G$ using coupling capacitances of equal magnitude. The sinusoidal injection locking signal at twice the oscillator frequency $f_0$ is applied across the injection capacitances. We implement a linear annealing schedule where the amplitude of the injection locking signal $V_{inj}$ is linearly ramped from zero to a maximum of 10V peak-to-peak over an annealing time $T_{anneal}$. This is followed by a phase readout time $T_{readout}$. Thus, the total computation time $T_{comp} = T_{anneal} + T_{readout}$. To experimentally investigate the efficacy of annealing, we vary the annealing time $T_{anneal}$ as shown schematically in Fig. 4(b). For all the cases, we keep $T_{readout}$ fixed at 100 oscillation cycles, while $T_{anneal}$ is varied from 0 (representing no anneal scenario) to 660 oscillation cycles.

Fig. 4(c) shows, for a single run, the evolution of the phases of the PTNOs for $T_{anneal} = 3.7\ ms$. The equivalent number of oscillations is calculated as $N_{osc} = 250\ cycles$. The temporal evolution of the Ising energy and the resultant cut set $C$ is shown in Fig. 4(d) for the case of no anneal and 250 cycle anneal. For the case of no anneal, the application of a high $V_{inj}$ immediately binarizes the phases of the PTNOs. Hence, the corresponding temporal evolution of the Ising energy shows a steep descent. However, as highlighted in Fig. 2(f), the high $V_{inj}$ results in a high dwell time that significantly reduces the temporal fluctuations in the oscillator phases. Thus, with very little freedom to escape the local minima, the network converges to a sub-optimal solution with a higher energy. On the other hand, when we linearly increase $V_{inj}$ over 250 cycles, the dwell time exponentially increases as highlighted in Fig. 2(f) and the temporal fluctuations in the oscillator phases gradually reduce. The network slowly performs energy minimization with more freedom to escape the local minima and converges to the optimal solution with a higher probability as shown in Fig. 4(d). Thus, we can perform classical annealing in our PTNO-CTDS by controlling the temporal fluctuation in the oscillator phases which is very similar to that of simulated annealing with a decaying noise (or temperature). To quantify the performance of the PTNO-CTDS for varying annealing conditions, we run the network multiple times to calculate the success probability for finding the Max-Cut on this graph instance. The success probability is defined as the ratio of the number of trials that returned the true ground-state energy to the total number of trials. To obtain the true ground-state energy for comparison, we run the same graph instance using the BiqMac solver that executes an exact algorithm (branch-and-bound) on a digital CPU (*28*) and using a QUBO software called qbsolv (*29*) developed by D-Wave. Fig. 4(e) shows the success probability increasing with the annealing cycles. The scenario of no anneal resulted in a success probability of 30% obtained experimentally while increasing $T_{anneal}$ to over 600 cycles resulted in a success probability of 96%. It is to be noted that the measured performance of our Ising solver is limited by the parasitics introduced by the experimental setup. One major contributor is the parasitic coupling capacitances $C_{C,W}$ arising from the breadboard. The presence of a large $C_{C,W}$ (estimated to be around 22pF in our experiment) introduces undesired coupling among the oscillators in addition to the intended coupling $C_C$ determined by the adjacency matrix and hence lowers the success probability (see supplementary section S8 for details). To compare with experimental results, we perform numerical simulations for an 8-oscillator network coupled using capacitances as shown in Fig. 4(a). We use a PPV-based framework with experimentally calibrated device and circuit parameters. The simulation details are delineated in the Methods section and supplementary section S2 and S8. We also introduce the same annealing scheme as our experiments. The success probability obtained from the simulations show very good agreement with our experimental results as shown in Fig. 4(e). Overall, the experimental and simulation results validate our proposed methodology of progressively obtaining better solutions through annealing. The excellent agreement of our numerical simulations with experimental results also enables us to use the simulation framework to predict the performance of the PTNO-CTDS for larger problem size.

**Scaling with Problem Size**
Next, we use the experimentally calibrated PPV-based numerical simulation framework to explore and benchmark the performance of our PTNO-CTDS-based Ising solver for solving Max-Cut with increasing problem size. It is to be noted that such analog computing using CTDS exhibits an inevitable challenge arising from parasitics and variability as we scale up to larger network size. As such, in our simulation framework, we incorporate the non-idealities such as interconnect/wire parasitics in terms of line-to-ground capacitance, line-to-line capacitance and frequency variability among the oscillators. The coupling capacitance $C_C$ is optimally chosen in our simulations such that the parasitic coupling capacitance remains an order or magnitude lower. This ensures that we obtain high success probability (see supplementary section S6 and S8). The simulation details are highlighted in the Method section and supplementary sections S6 and S7. We first start by investigating the performance of the Ising solver on a 100 node Mobius Ladder graph which is a regular cubic graph of degree 3 as shown in Fig. 5(a). We use a linear anneal scheme where the amplitude of the SHIL is increased over 500 cycles. Fig. 5(b) shows the temporal evolution of the oscillator phases governed by the process of energy minimization. Fig. 5(c) shows the decrease in the Ising energy accompanied by an increase in the cut size as the network evolves towards the ground state configuration. We run the simulation 100 times to calculate a success probability of 94%. Next, we extend the investigation to random cubic graphs of degree 3 for different problem sizes ranging from 10 to 100 nodes. Fig. 5(d) shows the success probability with increasing problem size. We compare simulation results for three different anneal cycles – 100, 500 and 1000. It is seen that, for 100 node problems, the success probability reduces to 15% for a short anneal time of 100 cycles. Increasing the anneal time to 500 and 1000 cycles boosts the success probability to 44% and 86%, respectively. Fig. 5(e) shows the success probability for dense Max-Cut problems (connectivity of 50%) as a function of problem size and different anneal cycles. For the 100-node graph, the success probability increases from 10% for a short anneal time of 100 cycles to 27% and 80% for 500 and 1000 cycles, respectively. Overall, we see that for increasing problem size, it becomes difficult to find the global optimal solution. However, reducing the reduction in success probability can be mitigated by increasing the number of cycles. Note that, here we report the success probability for obtaining the absolute ground state or the optimum Max-Cut value. Relaxing the solution accuracy to 99% or 95% of the optimum Max-Cut value will incur a much higher success probability. This means that the PTNO-CTDS is capable of finding near-optimal solutions with high probability of success, which has immense implication for tackling real-world industrial problems.

**Performance Evaluation of PTNO-based Ising Solver**
A key metric for benchmarking the performance of any Ising solver is the total computation time required to obtain at least one ground state solution. The total time to solution is calculated as $T_{solution} = T_{comp} \times N_{runs}$. The computation time is given by $T_{comp} = T_{anneal} + T_{readout}$, involving both the annealing and readout times. The annealing time is calculated as $T_{anneal} = \left(\frac{N}{N_b} T_{clock}\right) \times N_{cycles}$. Here, $N$ denotes the problem size, $N_b$ denotes the batch size for updating the spins and $T_{clock}$ is given by the operating frequency. The first term gives the time required to run one annealing cycle. To obtain a higher success probability, annealing has to be performed over a larger $N_{cycles}$. Finally, with increasing graph size, the success probability $P_{success}$ of a single run decreases exponentially, necessitating the solver to re-run the problem $N_{runs}$ times for ensuring a 99% cumulative success probability of obtaining at least one ground state solution. We calculate $N_{runs} = [\log(1 - 0.99)]/[\log(1 - P_{success})]$. Since our PTNO network operates in continuous-time in a dynamical fashion, it allows synchronized updates of all the oscillator phases. Thus, effectively we have a batch size of $N_b = N$. In other words, the time for a single annealing cycle is not bounded by the batch size, but rather by the operating frequency. In our Ising solver, the operating frequency of the PTNOs decrease with increasing problem size due to the impact of parasitics and coupling capacitance. To incorporate such effects, we perform a detailed SPICE circuit simulation to calculate the slowdown in the operating frequency of the oscillator network. By including the parasitic capacitances in the simulation, we estimate an operating frequency of around 500MHz for a small network size while the frequency decreases to 87MHz for a 100-node oscillator network. Note that the upper limit for the frequency will be bounded by the intrinsic switching capacitance of the VO$_2$ which is estimated to be around $41fF/um$ (*30*) (see supplementary section S6 for details). The number of annealing cycles required on the other hand is strictly determined by the energy minimization property of the solver. As shown in Figs. 5(d, e), we require $N_{cycles} = 1000$ to obtain high success probability for larger problem sizes.

Fig. 6(a) shows the total time-to-solution for dense Max-Cut problems with 50% connectivity as a function of the problem size $N$ follows an exponential nature ($ae^{bN}$) highlighting the NP-hard complexity of the problem. The different time-to-solution curves for different anneal times intersect each other for increasing problem size, thus highlighting a non-trivial dependence of $T_{solution}$ on $T_{anneal}$. Optimizing $T_{solution}$ reveals a tradeoff between the anneal time for a single run and the success probability. It is seen that shorter anneal time of 100 cycles is preferred for small problems up to 50 nodes where the success probability remains very high and is therefore insensitive to the anneal time. However, longer anneal of 1000 cycles is preferred for larger problems where the success probability dominates. Thus, the optimum anneal time required increases with the problem size.

We also investigate another key metric for benchmarking the performance - the energy-to-solution for solving such graph problems. For the PTNO-CTDS, the average power consumption for the main compute kernel, i.e. the coupled oscillator network, is estimated to be around $20\mu W$ per oscillator as obtained from our circuit simulations. Note that an additional energy overhead will arise from the peripheral readout circuit. As such, we propose a CMOS readout circuit consisting of an SR-latch, low pass filters and a digital comparator for reading out the phases of the oscillators (see supplementary section S9 for details). Simulating the readout circuit in SPICE reveals an average power of 5.51uW. The total consumed energy–to-solution is then estimated considering the total time-to-solution from Fig. 6(a). Fig. 6(b) shows the energy-to-solution for dense Max-Cut problems with 50% connectivity for increasing problem size and for different anneal schemes. Similar to time-to-solution, we see that shorter anneal times are preferred for small problems where the success probability remains close to unity and is therefore insensitive to the anneal time. However, longer anneal cycles are preferred for larger problems where the success probability dominates.

**Performance Comparison with Other Approaches**
Table 1 shows the performance of the PTNO-CTDS-based Ising solver compared with other approaches for solving 100-node random dense Max-Cut problems. We highlight the relevant metrices for comparison such as time-to-solution, energy-to-solution, power dissipated and energy-efficiency (calculated as solutions per second per Watt). For the comparative study, we include five different approaches - (a) well-known simulated annealing algorithm (*8*) running on an iMac computer with four 3.5 GHz Intel Core i5 processors, (b) a noisy mean-field annealing algorithm running on an NVIDIA GeForce GTX 1080 Ti GPU (*31*), (c) D-Wave's 2000Q quantum annealer containing 2,048 qubits (*16*), (d) coherent Ising machine (CIM) based on optical parametric oscillator with an FPGA feedback loop (*14*, *15*) and (e) a discrete-time memristor-based hybrid analog-digital accelerator implementing Hopfield neural network (mem-HNN) (*23*). We see an overall similar time-to-solution for both the PTNO-CTDS and the mem-HNN approach. This is because the mem-HNN uses a hybrid scheme of updating 10 nodes per clock cycle (clock frequency 500MHzz). Since their best time-to-solution scenario utilize only 50 anneal cycles to obtain the Max-Cut solution, the total anneal time gets lowered to $1\mu s$. However, with a success probability of 15% for 100 nodes, the mem-HNN needs to be re-run at least 25 times, resulting in an overall time-to-solution of $25\mu s$. On the other hand, the PTNO-CTDS operating at 87MHz shows an optimum scenario of utilizing 1000 anneal cycles or $10\mu s$ anneal time that allows us to achieve 80% success probability and needs to be re-run only 3 times to obtain the Max-Cut solution. This provides an overall time-to-solution of $30\mu s$. In terms of energy-to-solution, we see a 5x improvement over mem-HNN due to the low power dissipation of the PTNOs. Overall, we obtain an energy-efficiency (measured in terms of solution per second per watt) of $1.3 \times 10^7$ which is 5x higher compared to mem-HNN. Compared to D-Wave's 2000Q quantum annealer, we see a orders of magnitude improvement in time-to-solution and energy-to-solution for dense Max-Cut problems. Note that the significantly high time-to-solution for D-Wave arises due to the fact that the dense problems with more that 61 nodes are not embeddable in the current DW2Q machine (*16*). The huge energy penalty for D-Wave comes from their cryogenic cooling need that requires around 25kW of power. We expect this overhead from cooling to be constant with scaling up to thousands of nodes. Thus, overall the benefits that might be obtained from D-wave's potential use of quantum effects are negligible for the problems considered in this work. However, we acknowledge that quantum annealers in general might offer speed-ups for tasks such as simulating quantum processes. The coherent Ising machine (CIM) based on degenerate optical parametric oscillator (DOPO) requires kilometer long fiber cavity to accommodate the DOPO pulses. This incurs a cavity round trip time of microseconds and puts an upper limit on the time-to-solution. For 100-node dense Max-Cut problems, we see a 76x improvement in time-to-solution compared to CIM. This is because even though the CIM uses a pump repetition frequency of 1GHz, it incurs an additional overhead of around 2.5 due to the delays from the DAC/ADC feedback circuits, FPGA performing the coupling computation and stabilization of the feedback loop. Thus, the total round-trip time is around $2.5 \mu s \times N$, where N is the problem size. For a 100-node dense Max-Cut problem, the effective anneal time for CIM can be calculated as $250 \mu s$. For around 40% success probability, the total time-to-solution becomes around 2.3ms (*16*). In comparison, we obtain 80% success probability for an anneal time of $10 \mu s$, which yields a time-to-solution of $30 \mu s$. It must be noted that although the DOPOs in CIM operate in continuous-time, it is completely asynchronous in nature as one pulse is updated per cycle. Performing updates in batches will potentially be possible for CIM, but that will drastic increase the system complexity. However, despite of the lower time-to-solution and higher energy dissipation of CIM, it must be acknowledged that the CIM currently exhibits scalability up to 100k spins which is yet to be demonstrated by other technologies. The noisy mean field algorithm running on a GPU allows all the nodes to be updated synchronously. Compared to GPU running at 1GHz clock frequency, we get a 1.3x improvement in time-to-solution owing to a short annealing time of $12.3 \mu s$, similar to our Ising solver (*31*). However, the scaling of annealing time for GPU non-trivial for dense problems involving dense matrix-vector multiplications. As such, we may expect a quadratic dependence (*23*). In terms of energy-to-solution, we get a four-orders of magnitude compared to GPU. Finally, we outperform CPU which is running the conventional simulation annealing algorithm in terms of time-to-solution and energy-to-solution by orders of magnitude.

Overall, report a high energy-efficiency of $1.3 \times 10^7$ solutions/sec/Watt which exhibits a 5x improvement over the recently demonstrated memristor-based Hopfield neural network and several orders of magnitude improvement over other candidates such as CPU and GPU, D-Wave and CIM. Such a performance gain can be attributed to (a) inherent advantage provided by the CTDS approach that allows synchronized updates of all the oscillator phases, (b) short annealing times to obtain high success probability and (c) low-power dissipation of PTNOs. While the success probability obtained in this work is by utilizing a linear annealing scheme, we believe further improvement is possible in terms of exploring better annealing methodologies or modifying the energy function to avoid non-solution attractor states that trap the system in local minima (*32, 33*). Hybrid approaches can also be adopted to improve the quality of solution, such as augmenting the search of an Ising solver in the first phase with other metaheuristic local-neighborhood search algorithms such as Tabu search (*34*) in the second phase. While the concept of utilizing coupled oscillator-based networks for performing computation such as solving optimization problems have been recently explored (*35–37*), they involve bulky LC oscillators and ring oscillators with latch-based coupling. In comparison, we demonstrate a compact hardware using capacitively coupled one-transistor and one-resistor (1T-1R) PTNOs which provides marked area and energy benefit (see supplementary information section S10 for details). Finally, it is to be noted that, while we propose capacitive coupling through a programming transistor to achieve programmable coupling scheme for all the N*(N-1)/2 connections (see supplementary information section S6), in practice an all-to-all connected oscillator network may not be feasible for a large problem size due to effect of parasitics. However, there lies various avenue of research that are currently being undertaken to handle large-scale real-life problems. These involve decomposing large-scale problems into smaller Ising/QUBO problems that are might be tractable with practically achievable oscillator networks (*29*, *38*).

**Conclusion**
The notion of solving hard optimization problems using the continuous-time dynamics of a physical system reveals new avenues of exploration of dynamical systems for compute applications. There is much enthusiasm in building special purpose machines (or accelerators) for solving graph problems belonging to the NP-hard and NP-complete complexity class as part of a strong push towards a heterogenous compute platform. There is a rapidly growing demand to analyze and uncover hidden relationships between similar or diverse datasets in real-time and service applications such as customer analytics, risk and compliance management, recommendation engines, route optimization, fraud detection, asset allocation and risk management. We are witnessing a resurgence in building dedicated optimization processing units (such as Ising Hamiltonian solvers) that can complement general-purpose CPU and GPU. Specialized hardware or accelerators such as Ising solvers are gaining attraction in real-life as many relevant NP-hard and NP-complete problems can be reformulated into the problem of finding the ground state of an Ising model (*4*). Here, we showcase that exploiting the vast repertoire of emergent complex dynamics exhibited by CTDS enables us to design special purpose hardware that are most appropriate for solving computationally hard optimization problems belonging to the NP-hard or NP-complete complexity class. We believe that the immense benefit of such a CTDS hardware in terms of operating speed and energy dissipation comes from the inherent capability of the system to perform collective computing in a distributed and highly parallel fashion.

**Methods**
**Sample preparation**
10nm think Vanadium dioxide ($VO_2$) is grown on a substrate of (001) $TiO_2$ substrate using Veeco Gen10 molecular beam epitaxy (MBE) system. The widths of the two terminals are defined by dry etching with $CF_4$. The device length is defined by depositing Pd/Au metal contacts using electron beam evaporation. The fabricated $VO_2$ devices varied in length from 100nm to 1um with resulting insulator-to-metal transition threshold voltages ranging from 0.7V to 4V. All our experiments have been performed on device lengths of $200 nm$.

**Experimental setup**
Fig. 1(c) shows the schematic of a $VO_2$-based PTNO realized by connecting an n-channel MOSFET (ALD1103) transistor in series with the two-terminal $VO_2$ device. A $V_{DD}$ of 2V is applied and the amplitude of the relaxation oscillations is ~1.7V. A gate voltage $V_{GS}$ = 0.8V is applied to the series transistor that set the oscillation frequency $f_0 \approx 100\ kHz$. Note that when the PTNOs are capacitively coupled, due to loading effect the frequency gets reduced to around $60 - 70\ kHz$. A schematic of the full experimental setup for the PTNO-based Ising solver is shown in Fig. 1(d). Eight VO$_2$ devices placed in the Keithley 4200-SCS probe station were connected using multi-contact probes. The V$_{DD}$ and the analog gate voltages V$_{GS}$ of the 8 series transistors are applied using a multichannel analog voltage card connected to a computer. The injection locking signal was applied to the 8 oscillators across injection locking capacitances $C_{inj} = 20pF$ using an external voltage generator. $C_{inj}$ was realized using discrete off-chip capacitors connected on a breadboard. The output voltage waveforms of the oscillators were measured using a multichannel digital oscilloscope. The coupling among the oscillators was realized using discrete off-chip coupling capacitances $C_c = 56pF$ connected on the breadboard.

**Data-processing**

The output voltage waveform of the oscillators was measured using a multichannel digital oscilloscope and subsequently analyzed and processed in MATLAB on a digital computer. The phase of an oscillator is calculated with reference to a reference sinusoidal signal with the same frequency $f_0$ as the oscillator and half of the injection locking signal. The phase is defined as the time difference between the minima point of the discharging phase of the IMT oscillator and the minima of the sinusoidal signal divided by the time period of the oscillator. The measured oscillator phases $\theta$ were converted to Ising spins $\sigma$ using a discretization window in the phase space such that if $0^0 < \theta < 180^0$, $\sigma = 1$, else $\sigma = 0$. Subsequently, the Ising energy and the cut set are calculated. The final energy state of the oscillator network and the corresponding final cut set is calculated during the readout phase as mentioned earlier. We readout the oscillator phases over 100 oscillation cycles.

**Numerical Simulation of PTNO network**

The numerical simulations for our PTNO network are based on the dynamical system theory as explained in the Supplementary Information section S1. We use a PPV-based numerical simulation methodology (*35*). For obtaining the PPV function $\xi$, we use a SPICE compatible macro-model of the IMT nano-oscillator (*39*) to quantitatively match the dynamics of the oscillators under injection locking conditions, and perform cycle accurate time domain simulations of the PTNO using the Cadence Spectre circuit simulation framework (*40*). The details of PPV calculation are described in the supplementary section S3. The stochastic differential equations describing the PTNO network is numerically solved using the Euler-Maruyama method. In our simulations, we consider a scaled VO$_2$ device of 100nm length. The insulator-to-metal transition voltage V$_{IMT}$ is considered as 0.7V and a V$_{DD}$ = 1V is used in our simulations. The simulation parameters are listed in Table S1 in supplementary section S2. The intrinsic capacitance of VO$_2$ (device to ground) is taken as $41fF/um$ (*30*). We use an insulating resistance of 200kΩ and a metallic resistance of 15kΩ. For realizing the PTNO, the transistor in series with the VO$_2$ is designed using TSMC 28nm logic technology node such that the ON resistance matches closely with the metallic resistance of the VO$_2$. Since the charging of VO$_2$ happens through the metallic resistance while the discharging happens through the series transistor, it is important to match the two resistances in order to obtain a symmetrical voltage waveform and hence a symmetrical PPV without any undesired harmonics (see supplementary section S8 for further details). It is interesting to note that the that the success probability of our Ising solver for finding the Max-Cut is sensitive to the coupling strength. As such, we varied the strength of the capacitive coupling to find an optimal coupling value that can maximize the success probability (see supplementary information section S5 for details). For the rest of the simulations, we use a coupling capacitance of $C_C = 6fF$. The oscillator jitter noise is taken from our experiments as 0.5%, considering a Gaussian distribution in the time period. We also investigate the impact of frequency mismatch among the oscillators on the success probability (see supplementary information S7 for details). We consider a Gaussian distribution of the oscillator frequencies. For the rest of the simulations, we consider a frequency mismatch of 0.1%. We use an injection locking capacitance $C_{inj} = 1fF$ in the simulations. We additionally incorporate the line-to-ground and line-to-line parasitic capacitances as highlighted in supplementary section S6. Incorporating the parasitic capacitances, we perform detailed SPICE circuit simulation to estimate the operating frequency as the network size increases (see supplementary section S6 for details).

**Data availability**
The data that support the findings of this study are available from the corresponding author upon request.

**Code availability**
The custom simulation code written in Matlab and SPICE for this study are available from the corresponding author upon request.

**Acknowledgements**
This work was supported in part by ASCENT, one of six centers in JUMP, sponsored by DARPA and the Semiconductor Research Corporation (SRC).

**Author contributions**
S. Dutta and S. Datta conceived the idea. S. Dutta and A.K. performed the measurements, analyzed the data and performed simulations. A.K. fabricated the devices. H.P. and D.S. did MBE growth of the $VO_2$ samples. A.R. and Z.T. participated in useful discussions. S. Dutta, A.K., A.R., Z.T. and S. Datta participated in the writing of the manuscript.

**Competing interests**
The authors declare no competing interests.

# Figure1

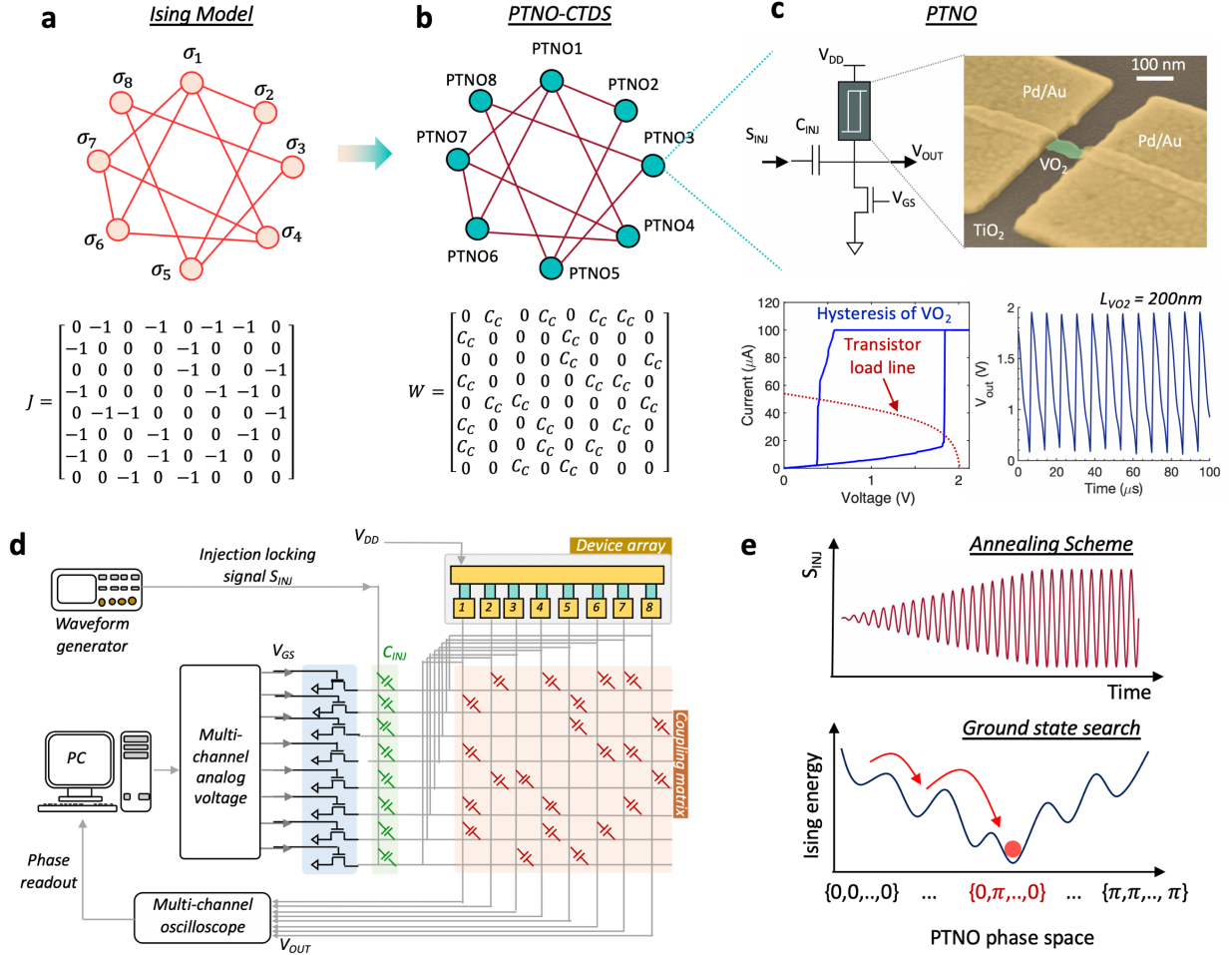

**Fig. 1. Overview of PTNO-based CTDS as an Ising Hamiltonian solver. (a, b)** The combinatorial optimization problem is reformulated in terms of the Ising Hamiltonian $H$ defined by the spin vector $\vec{\sigma}$, and the symmetric coupling matrix $J$ and mapped onto the Ising solver. Each Ising spin, representing a node in the graph is represented by an insulator-to-metal (IMT) phase-transition nano-oscillator (PTNO). The PTNOs are coupled to each other using passive elements such as capacitances. The coupling matrix W for the PTNO network is derived from the adjacency or coupling matrix $J$ of the Ising model. (c) Schematic of a PTNO consisting of a two-terminal phase-transition hysteretic device (VO$_2$) in series with a transistor. As the load line of the series transistor passes through the unstable hysteresis region of the VO$_2$ device, self-sustained oscillations are created. (d) Experimental setup of our PTNO-based CTDS. The main computing kernel comprises eight PTNOs connected using coupling capacitance following the coupling matrix W. The phenomenon of second harmonic injection locking (SHIL) is used to create bi-stable oscillator phases, emulating artificial Ising spin. (e) The inherent stochasticity present in the PTNO along with a novel technique of gradually reducing the temporal fluctuations in the oscillator phases by increasing the strength of the injection locking signal $S_{INJ}$ is utilized to perform classical annealing and obtain progressively better solutions.

**Figure2**

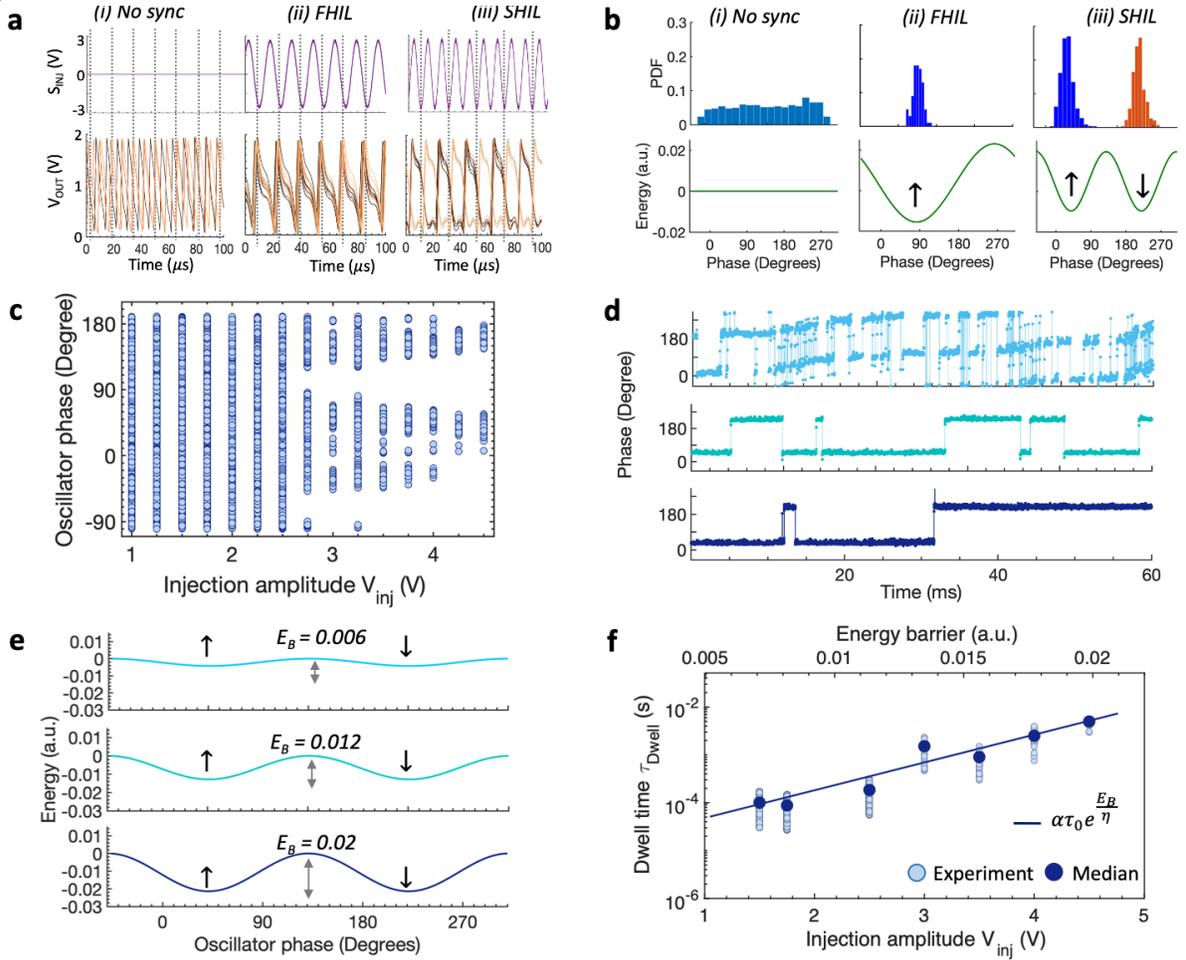

**Fig. 2. Creating artificial Ising spin using second harmonic injection locking (SHIL). (a)** Three different injection locking scenarios are shown. No synchronization case refers to free running oscillator with uniform of oscillator phase in the phase space. For first harmonic synchronization (FHIL), the oscillator remains injection locked with input signal at in-phase configuration. For second harmonic synchronization (SHIL), the oscillator phase gets binarized into in-phase and out-of-phase configuration with equal probability. **(b)** Measured distribution of the oscillator phases for the three scenarios along with the equivalent Ising energy for a single PTNO is shown depicting zero, one and two energy minima for stable phase locking, respectively. **(c)** Increasing the amplitude ($V_{inj}$) of SHIL forces the oscillator into bi-stable phase configuration. **(d)** The increase in $V_{inj}$ also reduces the temporal fluctuations between the two stable phases in the presence of noise. **(e)** Modulation of the energy landscape with $V_{inj}$. The three curves correspond to $V_{inj}$ = 1V, 3V and 5V. Increasing $V_{inj}$ increases the energy barrier between the two bi-stable energy minima states. This tightens the phase distribution as seen in (c). **(f)** The mean dwell time ($\tau_{Dwell}$) spent by an oscillator before hopping between the two phases is plotted against the applied injection locking amplitude and is found to follow a Arrhenius type law ($\alpha\tau_0 e^{E_B/\eta}$) with the inherent stochasticity ($\eta$) playing the role of temperature. $\tau_0 = \frac{1}{f_0}$ is the characteristic or attempt time (equal to the time period of the oscillator) and $\alpha$ is the fitting parameter.

**Figure 3**

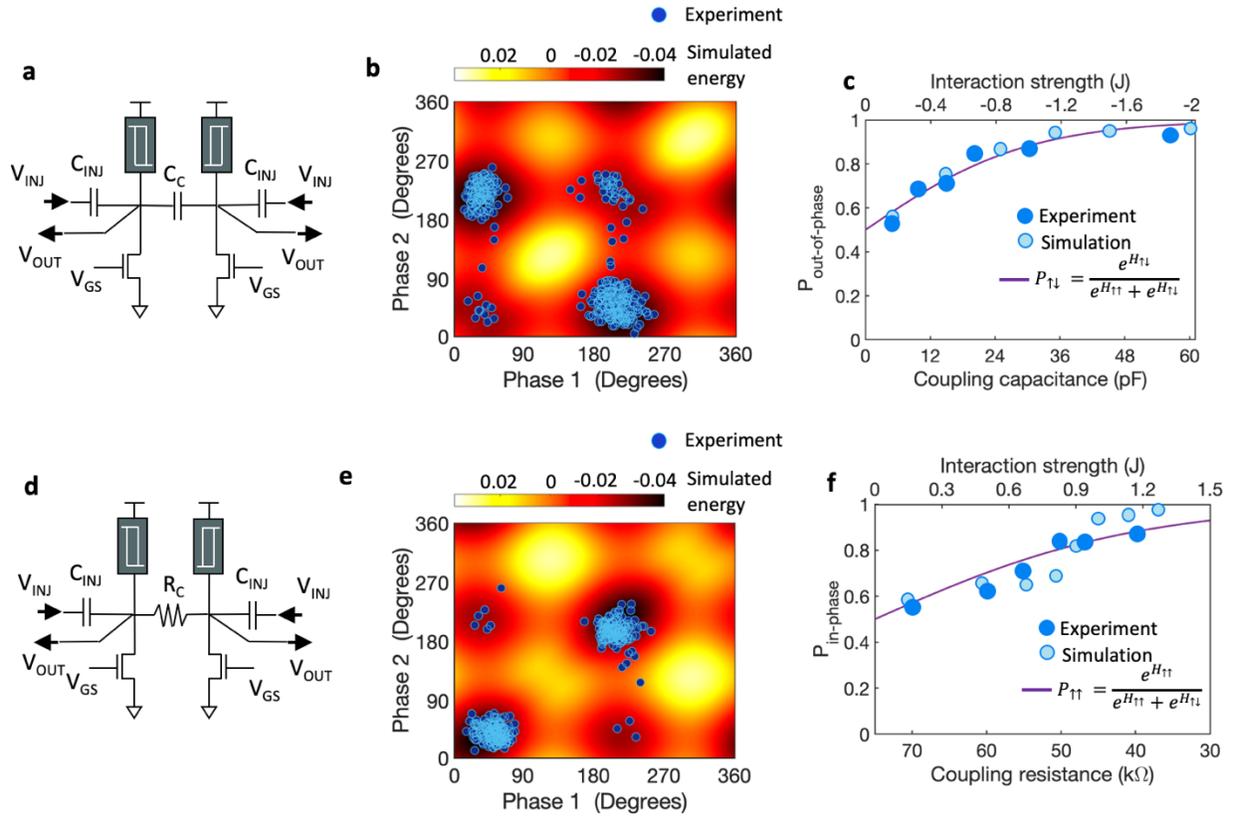

**Fig. 3. Replicating ferromagnetic and anti-ferromagnetic Ising interactions. (a)** Schematic of pairwise capacitively coupled PTNOs. **(b)** Measured distribution of the phases of the two PTNOs, highlighting the preference of the oscillators to remain in a stable out-of-phase configuration. The calculated energy landscape for capacitive coupling also highlights the presence of global energy minima corresponding to the out-of-phase configuration. **(c)** The measured probability of out-of-phase configuration as a function of varying capacitive coupling strength establishes the equivalent anti-ferromagnetic nature of interaction when compared to a 2-spin Ising model. **(d)** Schematic of pairwise resistively coupled PTNOs. **(e)** The measured phase distribution for resistive coupling along with the calculated energy landscape showing the preference for in-phase configuration. **(f)** The measured probability of in-phase configuration vs coupling strength highlights the ferromagnetic nature of interaction for resistively coupled oscillators.

**Figure 4**

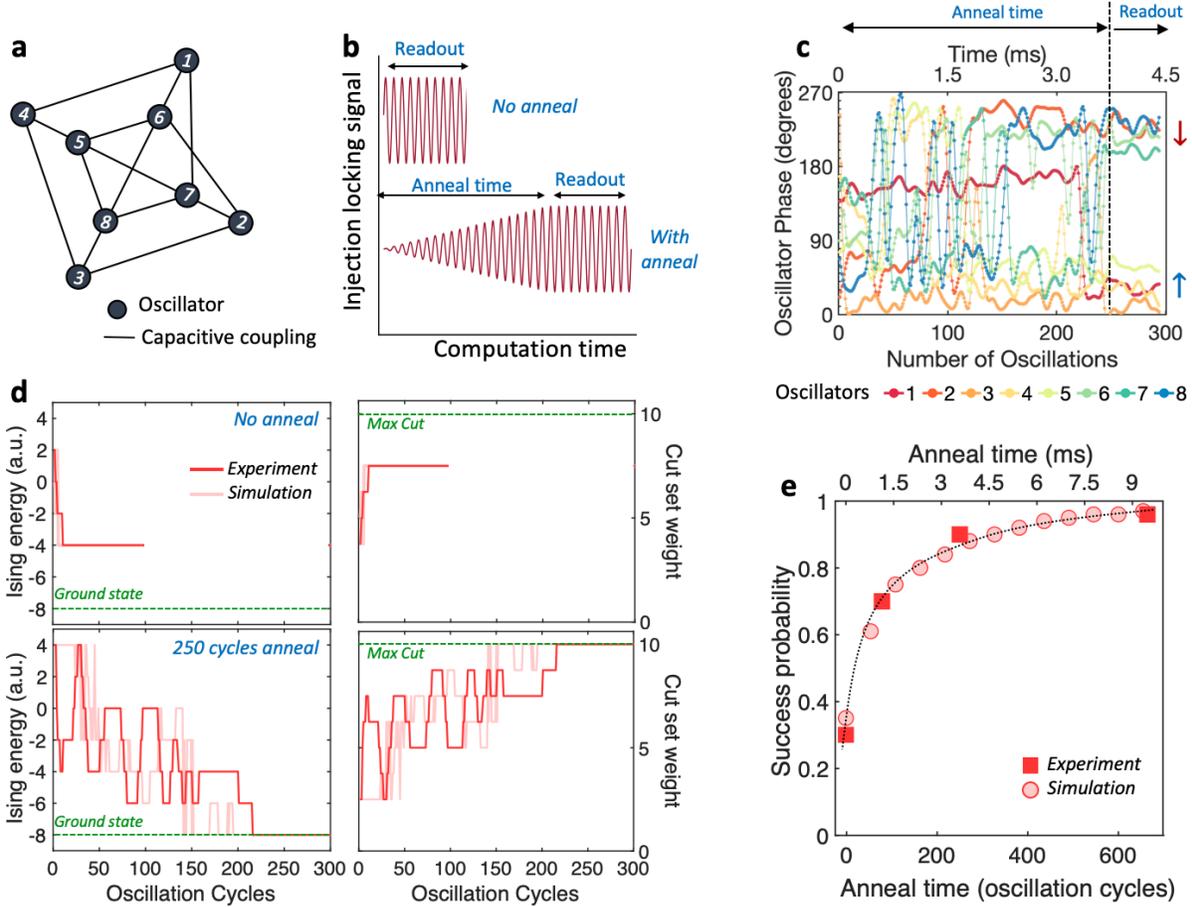

**Fig. 4. Experimental demonstration of Max-Cut and performance enhancement with annealing. (a)** An undirected and unweighted 8-node Mobius Ladder graph used for investigating the Max-Cut. The PTNOs are connected following the same adjacency (or connectivity) matrix of the graph using coupling capacitances. **(b)** Schematic of the annealing schedule used in the experiment. A sinusoidal injection locking signal at twice the oscillator frequency $f_0$ is applied with a linearly increasing amplitude over the annealing time $T_{anneal}$ that corresponds a linear annealing schedule. This is followed by a phase readout time $T_{readout}$. Thus, the total computation time $T_{comp} = T_{anneal} + T_{readout}$. The annealing time $T_{anneal}$ is varied from zero (corresponding to no anneal) to 660 oscillation cycles. **(c)** Evolution of the phases of the 8 oscillators, settling to either in-phase or out-of-phase with the injection locking signal. **(d)** The calculated temporal evolution of the Ising Hamiltonian shows an energy minimization accompanied by an increase in the graph cut size. For no annealing scheme, the network converges to a sub-optimal solution with a higher energy while annealing over 250 cycles allows the network to converge to the optimal solution with a lower Ising energy with a higher probability. Numerical simulations using same annealing schemes show very good agreement with our experimental results. **(e)** Experimental data and numerically simulation results for the success probability for different anneal times showing a steady increase from 30% with no anneal to 96% for over 600 cycles of linear anneal.

**Figure 5**

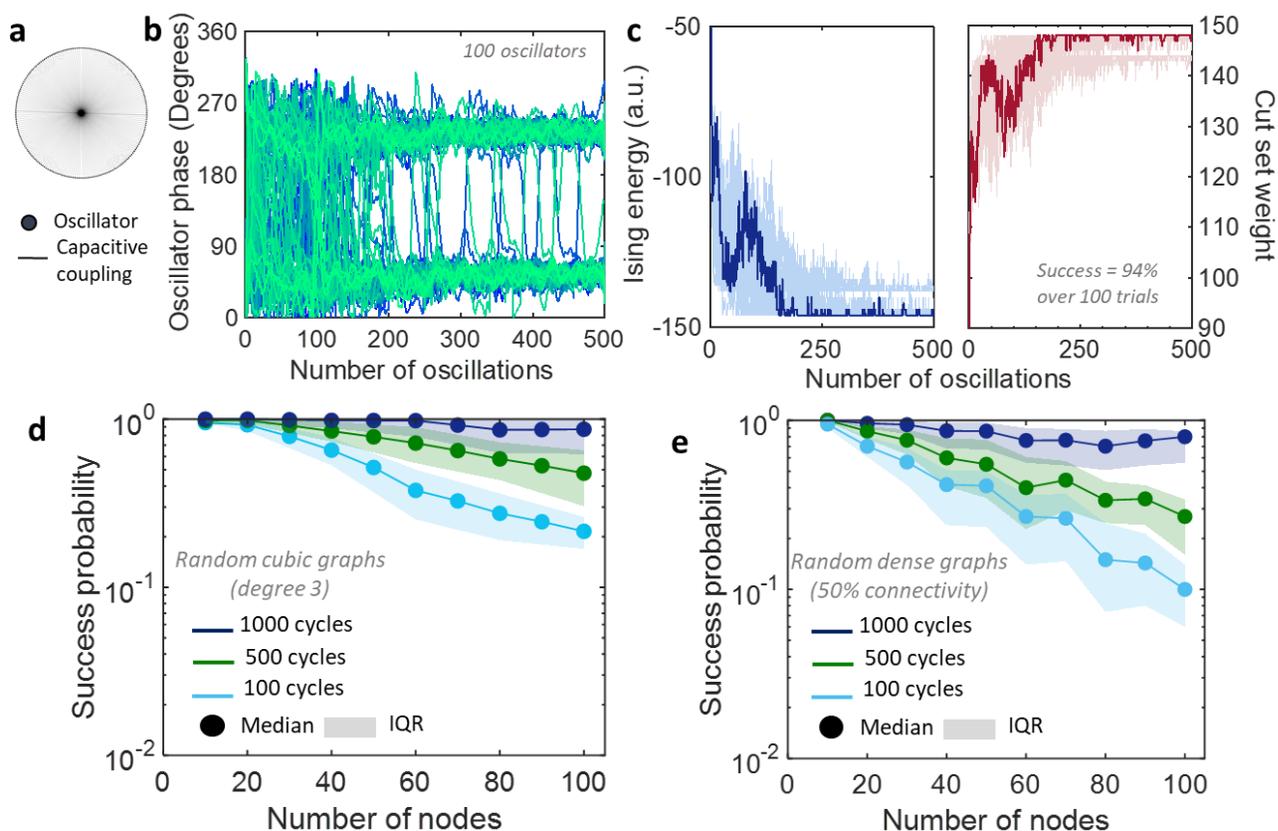

**Fig. 5. Scaling with problem size. (a)** Schematic of a 100 node Mobius ladder graph used for numerical simulation. **(b)** The evolution of the phases of the oscillators for a single run as a function of oscillation cycles. **(c)** Increase in the graph cut size accompanied by a decrease in the equivalent Ising energy as the system evolves towards the ground state configuration. The simulation was performed for 100 trials to calculate a success probability of 94%. **(d)** Success probability of finding the Max-Cut for random cubic graphs of varying size. We compare simulation results for three different anneal cycles – 100, 500 and 1000. **(e)** Success probability of solving dense Max-Cut problems with 50% connectivity for varying problem size. We consider three different anneal cycles – 100, 500 and 1000 for comparison.

**Figure 6**

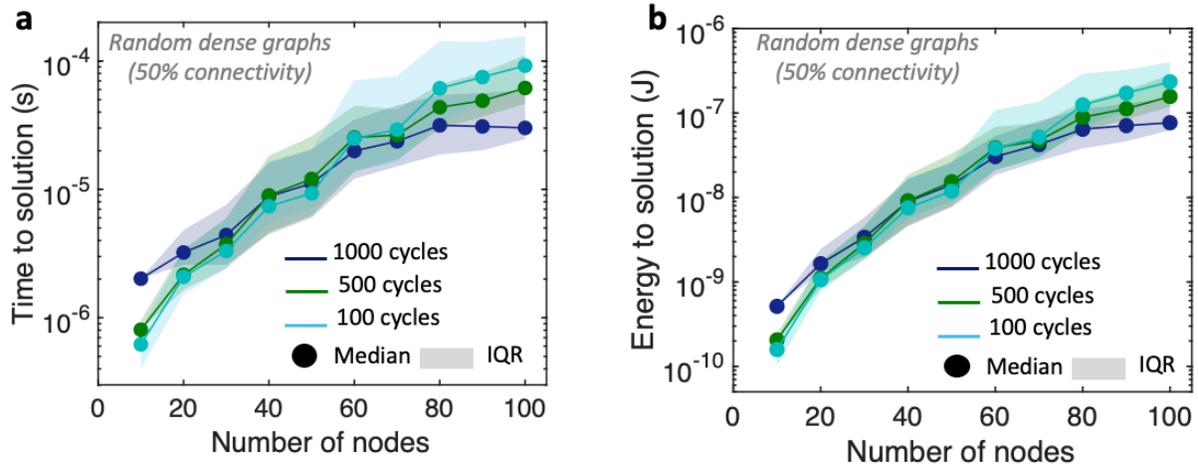

**Fig. 6. Performance evaluation of PTNO-based Ising solver. (a)** The total time-to-solution for solving dense Max-Cut problems with 50% connectivity for different anneal times and increasing problem size. A non-trivial dependence of time to solution on the anneal time is seen where shorter anneal time is preferred for smaller problems where the success probability remains insensitive to the anneal time. Longer anneal is preferred for larger problems where the success probability dominates. **(g)** Energy-to-solution for solving dense Max-Cut problems is plotted for different anneal times and for varying problem size showing a similar trade-off is seen.

**Table 1. Performance comparison between PTNO-CTDS-based Ising solver and other state-of-the-art approaches. Comparison done for solving Max-Cut on 100 nodes random cubic graphs.**

| | Simulated Annealing (CPU) | Noisy mean-field annealing (GPU) | D-Wave 2000Q | Coherent Ising Machine (CIM) | Memristor-based Hopfield network (mem-HNN) | Phase Transition Nano-oscillator (PTNO) |
|---|---|---|---|---|---|---|
| Represent spins | Spins | Spins | Qubits | Coherent light | - | Oscillator phases |
| Interaction | Ising interaction | Ising interaction | Flux storage | FPGA | - | Capacitance/Resistance |
| Update mechanism | Asynchronous | Synchronous | Synchronous | Asynchronous | Hybrid** | Synchronous |
| Connectivity | All-to-all | All-to-all | Sparse | All-to-all | All-to-all | All-to-all |
| Annealing scheme | Classical annealing | Classical annealing | Quantum annealing | Coherent computing | Modulate intrinsic noise | Classical annealing |
| Scaling with problem | $e^{-N}$ | $e^{-N}$ | $e^{-N^2}$ | $e^{-N}$ | $e^{-N}$ | $e^{-N}$ |
| Cryogenic cooling | No | No | Yes | No | No | No |
| Time to solution | 246ms | 40$\mu$s | >10$^4$s$^\$$ | 2.3ms | 25$\mu$s$^\#$ | 30$\mu$s$^{\#\#}$ |
| Power | 60W* | < 250W | 25kW | - | 16mW | 2.56mW |
| Energy to solution | 14.8J | < 10mJ | >250MJ | - | 400 nJ | 76.8nJ |
| Energy efficiency (Solutions/sec/Watt) | 6.7x10$^{-2}$ | > 100 | <4x10$^{-9}$ | - | 2.5x10$^6$ | 1.3x10$^7$ |

* SA algorithm implemented on a CPU using 4 Intel Core i5 processors each running at 3.5GHz
** Hybrid scheme updates 10 nodes at a time
$^\#$ Reported for 50 cycles (running sequentially)
$^{\#\#}$ Reported for 1000 cycles (running sequentially)
$^\$$ 10$^4$s for N = 55.